%% file: HIN-16-022_temp.tex
\pdfoutput=1

\documentclass[11pt,twoside,a4paper,cmspaper,final,collab]{cms-tdr}
\def\svnVersion{449483P}\def\cmsCernNoTag{CERN-EP-2017-220}\def\cmsCernDate{\today}\def\cmsMessage{Published in Physical Review Letters as \href{http://dx.doi.org/10.1103/PhysRevLett.120.092301}{\doi{10.1103/PhysRevLett.120.092301.}}}

\begin{document}\cmsNoteHeader{HIN-16-022}

\hyphenation{had-ron-i-za-tion}
\hyphenation{cal-or-i-me-ter}
\hyphenation{de-vices}
\RCS$HeadURL: svn+ssh://svn.cern.ch/reps/tdr2/papers/HIN-16-022/trunk/HIN-16-022.tex $
\RCS$Id: HIN-16-022.tex 440770 2018-01-09 15:56:48Z mguilbau $
\newlength\cmsFigWidth
\ifthenelse{\boolean{cms@external}}{\setlength\cmsFigWidth{0.98\columnwidth}}{\setlength\cmsFigWidth{0.8\textwidth}}
\ifthenelse{\boolean{cms@external}}{\providecommand{\cmsLeft}{top\xspace}}{\providecommand{\cmsLeft}{left\xspace}}
\ifthenelse{\boolean{cms@external}}{\providecommand{\cmsRight}{bottom\xspace}}{\providecommand{\cmsRight}{right\xspace}}
\newcommand {\roots}    {\ensuremath{\sqrt{s}}}
\newcommand {\rootsNN}  {\ensuremath{\sqrt{\smash[b]{s_{_{\mathrm{NN}}}}}}\xspace}
\newcommand {\deta}     {\ensuremath{\Delta\eta}}
\newcommand {\dphi}     {\ensuremath{\Delta\phi}}
\newcommand {\pp}    {\ensuremath{\Pp\Pp}\xspace}
\newcommand {\PbPb}  {\ensuremath{\mathrm{PbPb}}\xspace}
\newcommand {\pPb}  {\ensuremath{\Pp\text{Pb}}\xspace}
\newcommand {\AonA}  {\ensuremath{\mathrm{AA}}\xspace}
\newcommand{\noff}    {\ensuremath{N_\text{trk}^\text{offline}}\xspace}

\providecommand{\HIJING} {\textsc{hijing}\xspace}

\newcommand{ \dmean}[1]{\langle\langle #1 \rangle\rangle}

\cmsNoteHeader{HIN-16-022}
\title{Observation of correlated azimuthal anisotropy Fourier harmonics in \pp and \pPb collisions at the LHC}

\date{\today}

\abstract{
The azimuthal anisotropy Fourier coefficients ($v_n$) in 8.16\TeV \pPb data are extracted via long-range two-particle correlations as a function of event multiplicity
and compared to corresponding results in \pp and \PbPb collisions. Using a four-particle cumulant technique, $v_n$ correlations are
measured for the first time in \pp and \pPb collisions. The $v_2$ and $v_4$ coefficients are found to be positively correlated in all
collision systems. For high multiplicity \pPb collisions an anticorrelation of $v_2$ and $v_3$ is observed, with a similar correlation strength
as in \PbPb data at the same multiplicity. The new correlation results strengthen the case for a common origin of the collectivity
seen in \pPb and \PbPb collisions in the measured multiplicity range.
}

\hypersetup{%
pdfauthor={CMS Collaboration},%
pdftitle={Observation of correlated azimuthal anisotropy Fourier harmonics in pp and pPb collisions at the LHC},%
pdfsubject={CMS},%
pdfkeywords={CMS, physics, heavy ion, ridge, high multiplicity, small system, flow}}

\maketitle

Studies of multiparticle correlations provide important insights into the underlying
mechanism of particle production in high-energy collisions of both protons and nuclei.
A key feature of such correlations in ultrarelativistic nucleus-nucleus (\AonA) collisions
is the observation of a pronounced structure on the near side (relative azimuthal angle
$\abs{\dphi} \approx 0$) that extends over a large range in relative pseudorapidity
($\abs{\deta}$ up to 4 units or more). This feature, known as the ``ridge'', has been found over a wide range of \AonA  center-of-mass
energies and system sizes at both the 
RHIC~\cite{Adams:2005ph,Abelev:2009af,Alver:2008gk,Alver:2009id,Abelev:2009jv} 
and the LHC~\cite{Chatrchyan:2011eka,Chatrchyan:2012wg,Aamodt:2011by,ATLAS:2012at,CMS:2013bza}.
It is interpreted as arising primarily from the collective hydrodynamic flow of a
strongly interacting, expanding medium~\cite{Ollitrault:1992bk,Alver:2010gr}.
The azimuthal correlations of emitted particle pairs are frequently assessed
via their Fourier decomposition, $\rd{}N_\text{pair} / \rd\Delta\phi \propto 1 + \sum_{n} 2V_{n\Delta} \cos (n\Delta\phi)$, where
$V_{n\Delta}$ are the two-particle Fourier coefficients. The
single-particle azimuthal anisotropy Fourier coefficients $v_n$ can be extracted as $v_n = \sqrt{V_{n\Delta}}$ if factorization is assumed~\cite{Voloshin:1994mz}. The second ($v_2$)
and third ($v_3$) coefficients are known as elliptic and triangular flow, respectively~\cite{Alver:2010gr}.
In hydrodynamic models, $v_2$ and $v_3$ are directly related to the initial collision geometry and its fluctuations,
which influence the medium evolution~\cite{Alver:2010dn,Schenke:2010rr,Qiu:2011hf}. These Fourier components provide
insights into the fundamental transport properties of the medium.

The correlations of different orders of $v_n$ coefficients have been studied in \PbPb collisions
at the LHC using the event-shape engineering technique~\cite{Aad:2015lwa} and the symmetric cumulant (SC)
method~\cite{ALICE:2016kpq, Bilandzic:2013kga,PhysRevC.95.044902}. It is found that
the $v_2$ coefficient exhibits a negative correlation with the $v_3$ coefficient, while
the correlation is positive between the $v_2$ and $v_4$ coefficients, across the full \PbPb centrality range.
These correlations have been shown to be sensitive probes of initial-state fluctuations ($v_2$ vs. $v_3$)
and medium transport coefficients ($v_2$ vs. $v_4$)~\cite{Giacalone:2016afq,ALICE:2016kpq,PhysRevC.95.044902}.

Strong collective azimuthal final-state anisotropies have been observed in high-multiplicity \pp and \pPb collisions,
similar to those in \AonA  collisions~\cite{Khachatryan:2010gv,Aad:2015gqa,Khachatryan:2015lva,Khachatryan:2016txc,CMS:2012qk,alice:2012qe,atlas:2012fa,Aaij:2015qcq,Khachatryan:2014jra,ABELEV:2013wsa,Chatrchyan:2013nka,Aad:2014lta,Khachatryan:2015waa}.
The origin of collectivity in these small systems is still under debate, see for example Ref.~\cite{Dusling:2015gta}.
Measurements of the correlations between $v_n$ coefficients in small systems will provide
new insights on the origin and properties of the observed long-range collectivity. Quantitative
hydrodynamic predictions of azimuthal correlations in \pp and \pPb systems
still have large uncertainties, mainly due to limited knowledge of initial-state fluctuations of energy deposition at 
sub-nucleonic scales~\cite{Dusling:2015gta,Schlichting:2014ipa,Bozek:2016kpf}. Detailed modeling of
initial-state fluctuations in \pp and \pPb collisions~\cite{Welsh:2016siu} can be further constrained by the study of
$v_n$ coefficient correlations. For example, a positive correlation between $v_2$ and $v_3$
is predicted in \pp collisions over the full multiplicity range~\cite{Welsh:2016siu}, the opposite to what is observed in \PbPb collisions~\cite{ALICE:2016kpq}.
Measuring $v_n$ correlations in small colliding systems will help to understand if a common paradigm to describe collectivity in all hadronic systems can be found.

This Letter presents high precision measurements of anisotropy coefficients $v_4$ in \pp at $\roots = 13$\TeV, \pPb at $\rootsNN =8.16$\TeV, and \PbPb\ at $\rootsNN =5.02$\TeV
using data from the CMS experiment. The 8.16\TeV \pPb data provide access to higher multiplicities than previously experimentally accessible. The first measurement of 
correlations of different $v_n$ in 13\TeV \pp, 5.02 and 8.16\TeV \pPb, and 5.02\TeV \PbPb are also presented. The $v_n$ coefficients are 
extracted via long-range ($\abs{\deta}>2$) two-particle correlations as a function of charged particle multiplicity. The $v_n$ results are compared to 5.02\TeV \PbPb, 
as well as previously published ones in 13\TeV \pp~\cite{Khachatryan:2016txc} and 5.02\TeV \pPb~\cite{Khachatryan:2015waa} 
collisions. Correlations of $v_2$ vs. $v_3$ and $v_2$ vs. $v_4$ are measured using the four-particle SC method in \pp, \pPb, and \PbPb.

The central feature of the CMS apparatus is a superconducting solenoid of 6\unit{m}
internal diameter, providing a magnetic field of 3.8\unit{T}. Within the solenoid volume,
there are four primary subdetectors including a silicon
pixel and strip tracker detector, a lead tungstate crystal electromagnetic calorimeter
(ECAL), and a brass and scintillator hadron calorimeter (HCAL), each composed of a
barrel and two endcap sections. 
The silicon tracker measures charged particles within the range $\abs{\eta}< 2.5$. For charged
particles with transverse momentum $1 < \pt < 10\GeVc$ and $\abs{\eta} < 1.4$, the track resolutions are
typically 1.5\% in \pt and 25--90 (45--150)\mum in the transverse (longitudinal)
impact parameter~\cite{Chatrchyan:2014fea}. Iron and quartz-fiber Cherenkov hadron forward
(HF) calorimeters cover the range $2.9 < \abs{\eta} < 5.2$. A detailed description of the CMS detector 
can be found in Ref.~\cite{Chatrchyan:2008zzk}. The detailed Monte Carlo (MC) simulation of the CMS detector response is based on
\GEANTfour~\cite{GEANT4}.

The measurements presented in this Letter use data sets of 13\TeV \pp, 5.02 and 8.16\TeV \pPb, and 5.02\TeV \PbPb collisions with integrated luminosities of about 2\pbinv,
35\nbinv, 186\nbinv, and 1.2\mubinv, respectively. When measuring the $v_{n}$ coefficients in \pp and \pPb\ collisions, the same event may contain
multiple independent interactions (pileup), which constitutes a background for the analysis of high-multiplicity events.
The average number of collisions per bunch crossing in \pp and \pPb data varied between 0.1 to 1.3 and 0.1 to 0.25, respectively. A procedure similar 
to that described in Ref.~\cite{Chatrchyan:2013nka} is used for identifying and rejecting events with pileup. 
To further suppress this contamination in the 8.16\TeV \pPb data, where the pileup was more common, data from the highest luminosity periods are excluded,
resulting in an integrated luminosity of about 140\nbinv. The SC analysis is found to be insensitive to
pileup within the quoted experimental uncertainties and, therefore, the \pPb data sample of full recorded integrated luminosity is used.
The 5.02\TeV \PbPb data sample used for comparison is made of about 300 million peripheral (30--100\% central) events
where 100\% means no overlap between the two colliding nuclei~\cite{Miller:2007ri}. 
The same reconstruction algorithm is applied to the \pp, \pPb, and \PbPb events, in order to directly compare the three systems
at similar track multiplicities.

Minimum bias (MB) 8.16\TeV \pPb events are triggered by energy deposits in at least one of the two
HF calorimeters above a threshold of approximately 1\GeV and
the presence of at least one track with $\pt > 0.4$\GeVc\ in the pixel tracker.
In order to collect a large sample of high-multiplicity \pPb collisions, a dedicated
trigger was implemented using the CMS level-1 (L1) and high-level
trigger (HLT) systems. At L1, the total number of ECAL+HCAL energy towers above a threshold of 0.5\GeV in 
transverse energy (\ET) 
is required to be greater than a given threshold (120 and 150).
Track reconstruction is performed online as part of the HLT trigger with the identical reconstruction algorithm used offline~\cite{Chatrchyan:2014fea}.
For each event, the reconstructed vertex with the highest number of associated tracks is selected as the primary vertex.
The number of tracks with $\abs{\eta}<2.4$, $\pt > 0.4\GeVc$, and a distance of closest
approach less than 0.12\unit{cm} to the primary vertex is determined for each event and is required
to exceed a certain threshold to enrich the sample with high-multiplicity events.
In addition, events are also required to
contain a primary vertex within 15\unit{cm} of the nominal interaction point along the beam axis and 0.2\unit{cm} in the
transverse direction. The trigger, event reconstruction and selections used in 13\TeV \pp,
5.02\TeV \pPb or \PbPb collisions are similar to those in 8.16\TeV \pPb collisions, and are described
in previous correlation analyses~\cite{Khachatryan:2016txc,Chatrchyan:2013nka,Khachatryan:2010gv, Khachatryan:2016got}.

For all data sets analyzed, primary tracks, i.e. tracks that originate at the primary
vertex and satisfy the high-purity criteria of Ref.~\cite{Chatrchyan:2014fea},
are used to perform the correlation measurements as well as to define event categories
based on the charged-particle multiplicity (\noff). In addition, the impact parameter significance of the tracks with
respect to the primary vertex in the longitudinal and the transverse direction are required to be less than 3 standard deviations. 
The relative \pt uncertainty must be less than 10\%. To ensure high tracking efficiency, only tracks with
$\abs{\eta}<2.4$ and $\pt > 0.3\GeVc$ are used in this analysis~\cite{Chatrchyan:2014fea}. 

The \pp, \pPb, and \PbPb data are compared in classes of \noff, where \noff is the number of
primary tracks with $\abs{\eta}<2.4$ and $\pt >0.4$\GeVc. The event classes are the same as in Refs.~\cite{Khachatryan:2015lva,Khachatryan:2016txc}.

The analysis techniques for two-particle correlations, 
averaged over $0.3<\pt<3.0$\GeVc, 
are identical to those used in
Refs.~\cite{Chatrchyan:2011eka,Chatrchyan:2012wg,CMS:2012qk,Chatrchyan:2013nka,Khachatryan:2014jra,Khachatryan:2015lva,Khachatryan:2015waa}.
The results are compared to published 5.02\TeV \pPb~\cite{Khachatryan:2015lva} data.
The $v_4$ coefficient in \pp collisions at $\roots = 13\TeV$ is also
measured, while the $v_2$ and $v_3$ coefficients have been obtained from Ref.~\cite{Khachatryan:2016txc}.
The SC technique was first introduced by the ALICE
Collaboration~\cite{ALICE:2016kpq} and is based on four-particle correlations using cumulants.
The main difference between the standard cumulant calculation and SC lies in the fact that the
former is used to compute diagonal $v_{n}$ terms and the latter is used for correlations between different
coefficient orders. The framework for the calculation is the same as the one used in standard cumulant
analysis and is based on the generic code distributed by Bilandzic et al.~\cite{Bilandzic:2010jr}.

To study the correlation between the Fourier coefficients $n$ and $m$, one can build 2- and 4-particle correlators with:
\ifthenelse{\boolean{cms@external}}{
\begin{linenomath}
\begin{equation}
\begin{aligned}
\dmean{2}_{n} &\equiv \left<\left<\re^{i(n\phi_{1} - n\phi_{2})} \right>\right>\\
\dmean{4}_{n,m} &\equiv \left<\left< \re^{i(n\phi_{1} + m\phi_{2}-n\phi_{3}-m\phi_{4})} \right>\right>,
\end{aligned}
\label{4pCorrelationSingleEvent}
\end{equation}
\end{linenomath}
}{
\begin{linenomath}
\begin{equation}
\dmean{2}_{n} \equiv \left<\left<\re^{i(n\phi_{1} - n\phi_{2})} \right>\right>\qquad
\dmean{4}_{n,m} \equiv \left<\left< \re^{i(n\phi_{1} + m\phi_{2}-n\phi_{3}-m\phi_{4})} \right>\right>,
\label{4pCorrelationSingleEvent}
\end{equation}
\end{linenomath}
}
where $\left<\left<\ldots\right>\right>$ denotes the average correlations over all events. The final observable, the $SC$, is defined as follows:
\begin{linenomath}
\begin{equation}
\label{eq:Fig79}
SC(n,m) = \dmean{4}_{n,m} - \dmean{2}_{n} \, \dmean{2}_{m}.
\end{equation}
\end{linenomath}

Expressed as a function of $v_n$, the symmetric cumulant $SC(n,m)$ measures correlations of Fourier coefficients
between the order of $m$ and $n$:
\begin{linenomath}
\begin{equation}
\label{eq:Fig80}
SC(n,m) = \left<v^{2}_{n}v^{2}_{m}\right> - \left<v^{2}_{n}\right>\left<v^{2}_{m}\right>,
\end{equation}
\end{linenomath}
where $\left<\ldots\right>$ denotes the average over all events. In this analysis, we compute a SC for events belonging to the same event multiplicity class (\noff) and
with the same number of tracks entering in the calculation (i.e., $N_\text{trk}^\text{ref}$ with $0.3 < \pt < 3.0$\GeVc).
Then, the different SCs are combined into larger bins by using the total number of 4-particle combinations as a weight, i.e.,
in an event with track multiplicity $M$, this weight equals $M(M-1)(M-2)(M-3)$.
This weighting procedure is necessary to reduce the impact of multiplicity fluctuations, which are particularly
relevant at low multiplicity~\cite{Khachatryan:2015lva,Khachatryan:2016txc}.

The systematic uncertainties of the experimental procedure are evaluated as a function of \noff by varying the conditions in
extracting $v_{n}$ coefficients and SCs for both 8.16\TeV \pPb and 5.02\TeV \PbPb samples.
For 13\TeV \pp and 5.02\TeV \pPb, the systematic uncertainties are taken from Refs.~\cite{Khachatryan:2016txc,Khachatryan:2015waa}.
Systematic uncertainties due to tracking inefficiency and misreconstructed track rate are studied by varying the track quality requirements.
The selection thresholds on the significance of the transverse and longitudinal track impact parameter divided by their uncertainties are varied from 2 to 5 standard deviations.
In addition, the relative \pt uncertainty is varied from 5\% to 10\%. The resulting systematic
uncertainty is found to be 1--2\% for $v_{n}$ and SCs depending on multiplicity in both colliding systems.
The sensitivity of the results to the primary vertex position along the beam axis ($z_\text{vtx}$) is quantified by comparing events with
different $z_\text{vtx}$ locations from $-$15 to +15\unit{cm}. The magnitude of this systematic effect is estimated to
be 1--2\%, depending on multiplicity, and is independent of colliding system and method ($v_n$ or SC).
For the 8.16\TeV \pPb sample, two additional sources of systematic uncertainties are investigated.
To study potential trigger biases, a comparison to high-multiplicity \pPb data for a given multiplicity range that have been collected by a lower threshold trigger
with 100\% efficiency is performed. This uncertainty is found
to be less than 1\%. The possible contamination by residual pileup interactions is also studied by varying the pileup
selection of events in the performed analysis, from no pileup rejection at all to selecting events with only one
reconstructed vertex. For $v_{n}$ results, this effect is more important at high multiplicities (3\%) than at
low ones (0.1\%). For the SC method, 
it is independent
of multiplicity and estimated to be 1\%. The total systematic uncertainty is estimated to be 1.7--4.1\% for 
$v_n$ depending on multiplicity and 1.8\% for SCs.

\begin{figure*}[thb]
\centering\includegraphics[width=\textwidth]{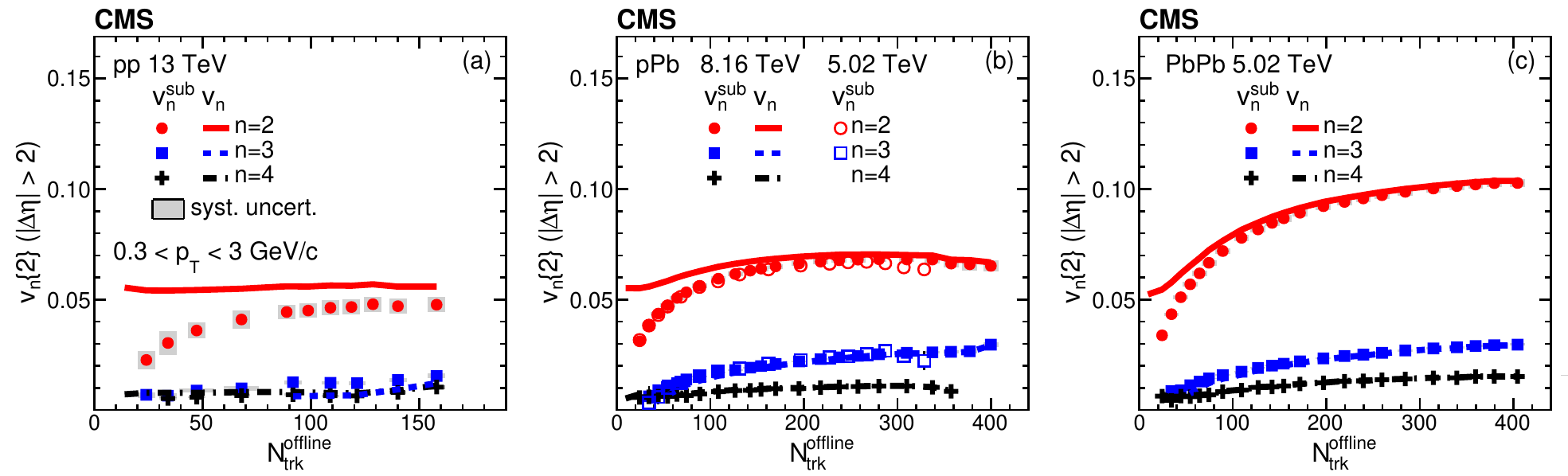}
    \caption{ The $v_2$, $v_3$~\cite{Khachatryan:2016txc}, and $v_4$ coefficients from long-range two-particle correlations
    as a function of \noff in 13\TeV \pp (a), 5.02\TeV~\cite{Chatrchyan:2013nka} and 8.16\TeV \pPb (b),
    and 5.02\TeV \PbPb collisions (c). The results corrected by low-multiplicity subtraction are denoted as $v_{n}^\text{sub}$. The
    lines show the $v_{n}$ results before subtraction of low-multiplicity correlations. The gray boxes represent systematic uncertainties.
    }
    \label{fig:vn2_all}
\end{figure*}

Measurements of $v_2$, $v_3$, and $v_4$ coefficients for $0.3<\pt<3$\GeVc extracted from
long-range two-particle correlations are shown in Fig.~\ref{fig:vn2_all}, as a
function of multiplicity in 13\TeV \pp, 5.02 and 8.16\TeV \pPb, and 5.02\TeV \PbPb collisions. 
The contribution to $v_n$ coefficients from back-to-back jet correlations are corrected
by subtracting correlations from very low-multiplicity events ($v_{n}^\text{sub}$), as done in
Refs.~\cite{Khachatryan:2016txc,Chatrchyan:2013nka}. The $v_{n}$ results before subtraction
are also shown as lines in Fig.~\ref{fig:vn2_all}. For $\noff> 200$, the low-multiplicity subtraction
has very small effect in \pPb and \PbPb collisions. At low multiplicity, this correction plays a larger role, in
particular for \pp collisions where dijet correlations are expected to be the main source of correlations.

By comparison with 5.02\TeV \pPb data, the new 8.16\TeV \pPb\ results extend the measurements of $v_{n}$ coefficients to a 
higher-multiplicity region, due to the higher collision energy and integrated luminosity.
The $v_2$ coefficient increases with \noff, saturating for $\noff > 200$.
Finite $v_4$, which are about 50\% smaller than the $v_3$ coefficients for $\noff > 100$, are also observed in all three systems.

\begin{figure*}[hbt]
\centering
\includegraphics[width=\textwidth]{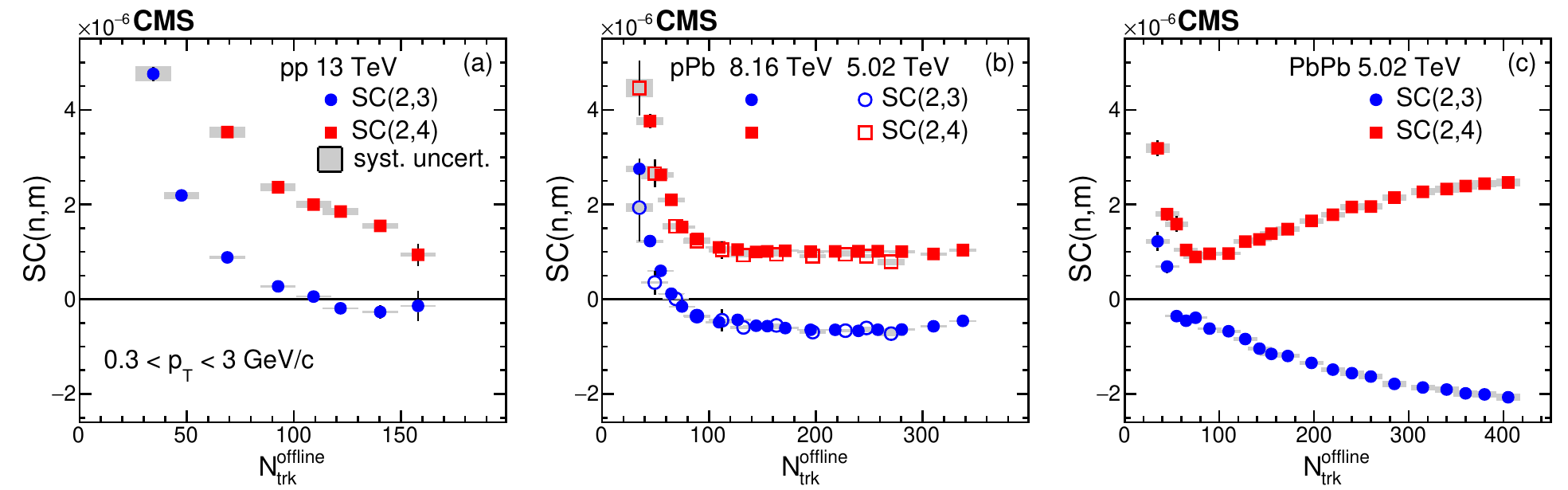}
    \caption{ The SCs for the second and third coefficient (red points) and the second and fourth coefficient (blue points) as a function of \noff
    in 13\TeV \pp (a), 5.02\TeV and 8.16\TeV \pPb (b), and 5.02\TeV \PbPb collisions (c).
    The gray boxes represent systematic uncertainties.
    }
    \label{fig:sc_all}
\end{figure*}

Measurements of symmetric cumulants $SC(2,3)$ and $SC(2,4)$ for $0.3<\pt<3$\GeVc from four-particle correlations are shown in Fig.~\ref{fig:sc_all}, as a
function of multiplicity in 13\TeV \pp, 5.02 and 8.16\TeV \pPb, and 5.02\TeV \PbPb, to further study the correlations of
different $v_n$ coefficients. 

In \pp collisions, both $SC(2,3)$ and $SC(2,4)$ decrease as \noff increases.
The $SC(2,4)$ values always remain positive, while there is an indication of a transition
to negative values for $SC(2,3)$ at $\noff>110$ but the measurement is not precise enough to draw a firm conclusion.
For \pPb and \PbPb data at sufficiently high multiplicities (e.g., $\noff>60$),
clear negative values of $SC(2,3)$ are observed, while $SC(2,4)$ values are positive.
The \PbPb data are consistent with results reported at $\rootsNN = 2.76\TeV$~\cite{ALICE:2016kpq}.

In hydrodynamic models, correlations of $v_2$ and $v_3$ can be directly related to
the initial eccentricity correlations~\cite{Giacalone:2016afq,ALICE:2016kpq,PhysRevC.95.044902}.
Theoretical studies of $v_n$ correlations in small colliding systems were
performed based on purely eccentricity correlations~\cite{Welsh:2016siu}.
An anticorrelation of $v_2$ and $v_3$ in \pPb collisions has been predicted
at high multiplicities~\cite{Welsh:2016siu}, which is consistent with the experimental observation.
A positive correlation of $v_2$ and $v_3$ is predicted over the full multiplicity range in \pp collisions~\cite{Welsh:2016siu},
while a hint of anticorrelation is seen in the data at high multiplicity.
However, larger \pp data samples are needed to draw a definitive conclusion.
At low \noff ranges ($\noff< 100$) for all three systems,
both $SC(2,3)$ and $SC(2,4)$ have positive values, which increase as \noff decreases.
It should be noted that, in the low multiplicity region, short-range few-body correlations such as jets
are likely to have a dominant contribution, which need to be properly accounted for
before comparing to models of long-range collective correlations. Indeed, the jet contribution at low \noff might be
different in \pp, \pPb and \PbPb and lead to slightly different behaviors of the SCs in this multiplicity range as observed
in the data. Finally, calculations from initial state gluon correlations in the color-glass condensate framework
have also been shown to capture the signs of the $v_n$ correlation data~\cite{Dusling:2017dqg,Dusling:2017aot},
although it remains to be seen if the magnitude of correlations in the measured multiplicity
region can be quantitatively reproduced. Recently, new methods have been proposed to suppress the contribution from jets down 
to low multiplicities by introducing sub-events in the cumulant calculation \cite{DiFrancesco:2016srj,Jia:2017hbm}. Future studies 
using these methods will be of high interest to better understand the short-range correlation contribution to correlation 
measurements at low multiplicity.

\begin{figure}[hbt]
\centering
\includegraphics[width=\cmsFigWidth]{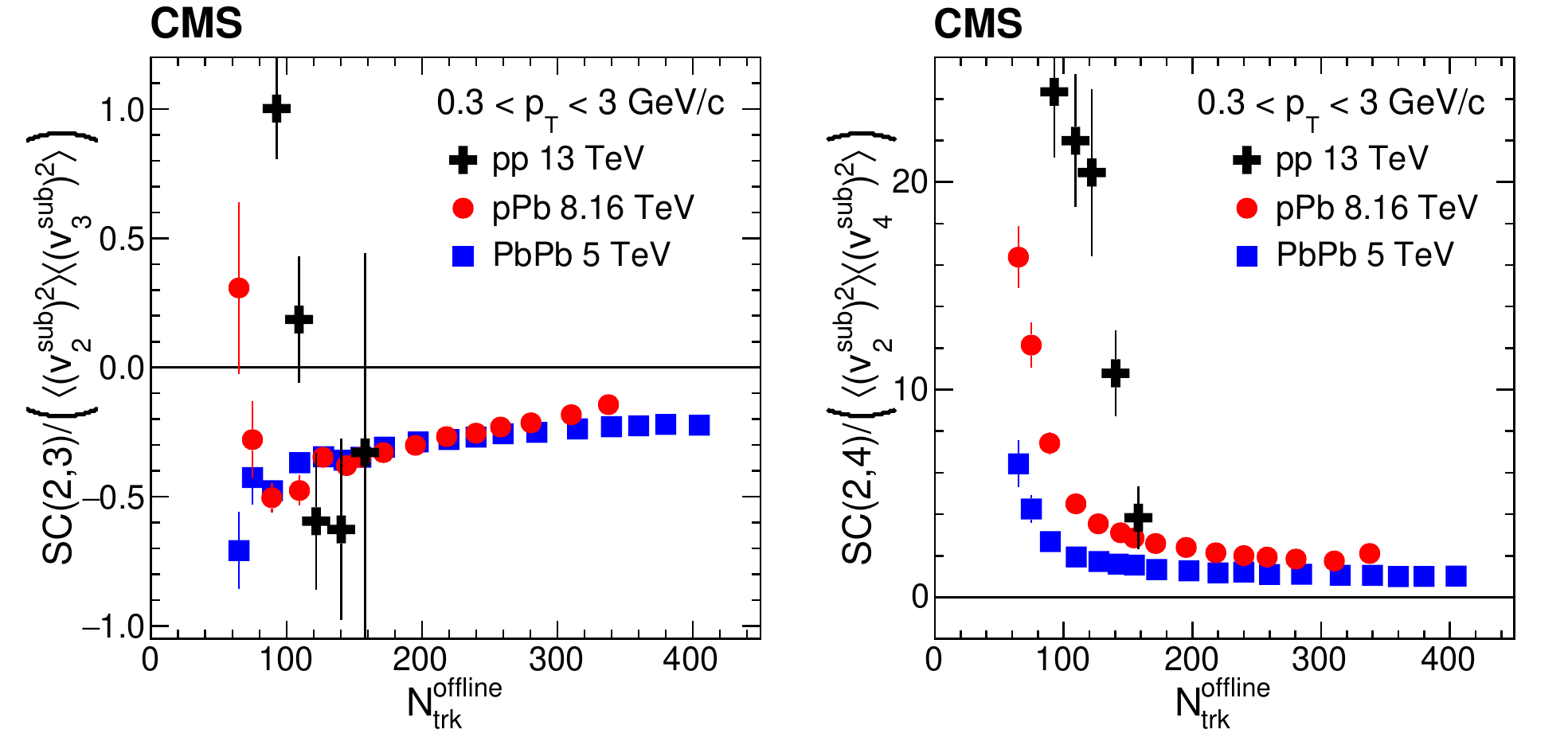}
    \caption{The SCs for the second and third coefficients (left) and the second and fourth coefficients (right) normalized by
    $\left<(v_2^{\text{sub}})^{2}\right> \, \left<(v_3^{\text{sub}})^{2}\right>$ and $\left<(v_2^{\text{sub}})^{2}\right> \, \left<(v_4^{\text{sub}})^{2}\right>$ from two-particle
    correlations. The results are shown as a function of \noff in 13\TeV \pp, 8.16\TeV \pPb, and 5.02\TeV \PbPb collisions.
    }
    \label{fig:sc_norm}
\end{figure}

The absolute magnitudes of $SC(2,3)$ and $SC(2,4)$ are found to be larger
in \PbPb than in \pPb system at high multiplicities. This may be related to the different magnitude of $v_n$ coefficients as indicated
in Fig.~\ref{fig:vn2_all}. To investigate the intrinsic correlation between $v_n$ coefficients and compare across different collision systems in a
more quantitative way, $SC(2,3)$ and $SC(2,4)$ are normalized by
$\left<(v^{\text{sub}}_{2})^{2}\right>\left<(v^{\text{sub}}_{3})^{2}\right>$ and $\left<(v^{\text{sub}}_{2})^{2}\right>\left<(v^{\text{sub}}_{4})^{2}\right>$,
respectively, based on the $v_n$ values from two-particle correlations in Fig.~\ref{fig:vn2_all}. As the two-particle correlation $v_{n}^{sub}$ with a 
rapidity gap is used for the normalization, the results might be affected by the event-plane decorrelation measured in Ref.~\cite{Khachatryan:2015oea} 
at the level of a few percent. Nevertheless, all systems would be affected consistently such that the conclusions from the results would not be modified. 
In addition, the short-range correlation contribution is suppressed with different approaches in the numerator (SC) and the denominator 
($\left<v_{n}^{2}\right>\left<v_{m}^{2}\right>$). The impact of the short-range correlation was investigated by using the unsubtracted $v_n$ for the 
normalization. As expected, at high multiplicity, the results remain unchanged. The resulting normalized SCs in all three colliding systems are shown in Fig.~\ref{fig:sc_norm}.

The normalized $SC(2,3)$ values are found to be very similar between \pPb and \PbPb systems at high multiplicities. Together with the $v_n$ results in Fig.~\ref{fig:vn2_all},
these measurements strongly suggest a unified paradigm to explain collective behavior observed
in large and small hadronic collisions.
In the context of hydrodynamic models, the $SC(2,3)$ data in \pPb and \PbPb collisions suggest
similar fluctuations of initial-state energy density of the collective medium~\cite{ALICE:2016kpq}. This common behavior may even apply to
\pp collisions for $\noff>120$, where $SC(2,3)$ tends to converge to a unified value for all three systems,
although statistical uncertainties are still too large to draw a firm conclusion.
The $SC(2,4)$, on the other hand, shows a clear dependence on the system size with a larger value for smaller systems.
The observed difference between $SC(2,4)$ values in \pPb and \PbPb collisions may
point to a different contribution of initial-state fluctuations or transport properties of
the medium such as the shear viscosity to entropy ratio~\cite{ALICE:2016kpq}.
Further calculations of $SC(2,3)$ and $SC(2,4)$ with full hydrodynamic evolution would be needed for a
quantitative comparison to the small system data.

In summary, the first measurements of azimuthal anisotropy Fourier coefficients and correlations
of different coefficients in 8.16\TeV \pPb collisions are presented based on
data collected by the CMS experiment at the LHC. The $v_2$, $v_3$, and $v_4$ Fourier coefficients are extracted from
long-range two-particle correlations in classes of event multiplicity, and are found to be
consistent with 5.02\TeV \pPb data.
The \pPb results are compared to those in 13\TeV \pp and 5.02\TeV \PbPb.
Using a four-particle cumulant technique, correlations of different coefficient orders
are obtained, where a negative (positive) correlation is observed between $v_2$ and $v_3$ ($v_4$) in \pPb collisions. This behavior is similar to what is observed in the \PbPb system, where the result is attributed to the hydrodynamic flow of a
strongly interacting medium. Normalized correlation coefficients for $v_2$ and $v_3$
are found to be quantitatively similar between \pPb and \PbPb, while for $v_2$ and $v_4$
the results are larger in \pPb than in \PbPb.
The corresponding result in \pp collisions shows a similar
trend at high multiplicity but the statistical uncertainties are too large to make a
quantitative statement. The results presented in this Letter provide
further evidence of a similar origin of collectivity observed in small and large
hadronic systems and impose constraints on theoretical
model calculations.

\begin{acknowledgments}
We congratulate our colleagues in the CERN accelerator departments for the excellent performance of the LHC and thank the technical and administrative staffs at CERN and at other CMS institutes for their contributions to the success of the CMS effort. In addition, we gratefully acknowledge the computing centers and personnel of the Worldwide LHC Computing Grid for delivering so effectively the computing infrastructure essential to our analyses. Finally, we acknowledge the enduring support for the construction and operation of the LHC and the CMS detector provided by the following funding agencies: BMWFW and FWF (Austria); FNRS and FWO (Belgium); CNPq, CAPES, FAPERJ, and FAPESP (Brazil); MES (Bulgaria); CERN; CAS, MoST, and NSFC (China); COLCIENCIAS (Colombia); MSES and CSF (Croatia); RPF (Cyprus); SENESCYT (Ecuador); MoER, ERC IUT, and ERDF (Estonia); Academy of Finland, MEC, and HIP (Finland); CEA and CNRS/IN2P3 (France); BMBF, DFG, and HGF (Germany); GSRT (Greece); OTKA and NIH (Hungary); DAE and DST (India); IPM (Iran); SFI (Ireland); INFN (Italy); MSIP and NRF (Republic of Korea); LAS (Lithuania); MOE and UM (Malaysia); BUAP, CINVESTAV, CONACYT, LNS, SEP, and UASLP-FAI (Mexico); MBIE (New Zealand); PAEC (Pakistan); MSHE and NSC (Poland); FCT (Portugal); JINR (Dubna); MON, RosAtom, RAS, RFBR and RAEP (Russia); MESTD (Serbia); SEIDI, CPAN, PCTI and FEDER (Spain); Swiss Funding Agencies (Switzerland); MST (Taipei); ThEPCenter, IPST, STAR, and NSTDA (Thailand); TUBITAK and TAEK (Turkey); NASU and SFFR (Ukraine); STFC (United Kingdom); DOE and NSF (USA).
\end{acknowledgments}
\bibliography{auto_generated}

\cleardoublepage \appendix\section{The CMS Collaboration \label{app:collab}}\begin{sloppypar}\hyphenpenalty=5000\widowpenalty=500\clubpenalty=5000\input{HIN-16-022-authorlist.tex}\end{sloppypar}
\end{document}

%% file: HIN-16-022-authorlist.tex
\textbf{Yerevan Physics Institute,  Yerevan,  Armenia}\\*[0pt]
A.M.~Sirunyan, A.~Tumasyan
\vskip\cmsinstskip
\textbf{Institut f\"{u}r Hochenergiephysik,  Wien,  Austria}\\*[0pt]
W.~Adam, F.~Ambrogi, E.~Asilar, T.~Bergauer, J.~Brandstetter, E.~Brondolin, M.~Dragicevic, J.~Er\"{o}, M.~Flechl, M.~Friedl, R.~Fr\"{u}hwirth\cmsAuthorMark{1}, V.M.~Ghete, J.~Grossmann, J.~Hrubec, M.~Jeitler\cmsAuthorMark{1}, A.~K\"{o}nig, N.~Krammer, I.~Kr\"{a}tschmer, D.~Liko, T.~Madlener, I.~Mikulec, E.~Pree, D.~Rabady, N.~Rad, H.~Rohringer, J.~Schieck\cmsAuthorMark{1}, R.~Sch\"{o}fbeck, M.~Spanring, D.~Spitzbart, W.~Waltenberger, J.~Wittmann, C.-E.~Wulz\cmsAuthorMark{1}, M.~Zarucki
% ...
\vskip\cmsinstskip
% ...
1:~~Also at Vienna University of Technology, Vienna, Austria\\
% ...


%% file: HIN-16-022_temp.bbl
\providecommand{\href}[2]{#2}\begingroup\raggedright\begin{thebibliography}{10}%
\makeatletter
\providecommand{\hrefCMSnoop }[0]{\@secondoftwo}%
\makeatother
\providecommand{\doi}{\texttt{doi:}\begingroup \urlstyle{tt}\Url}

\bibitem{Adams:2005ph}
\hrefCMSnoop {}{{STAR} Collaboration, ``{Distributions of charged hadrons
  associated with high transverse momentum particles in pp and Au+Au collisions
  at \rootsNN\ = 200\GeV}'',} \textit{ Phys. Rev. Lett.} \textbf{ 95} (2005)
  152301,
  \href{http://dx.doi.org/10.1103/PhysRevLett.95.152301}{\doi{10.1103/PhysRevLett.95.152301}},
\href{http://www.arXiv.org/abs/nucl-ex/0501016}{\texttt{arXiv:nucl-ex/0501016}}.

\bibitem{Abelev:2009af}
\hrefCMSnoop {}{{STAR} Collaboration, ``{Long range rapidity correlations and
  jet production in high energy nuclear collisions}'',} \textit{ Phys. Rev. C}
  \textbf{ 80} (2009) 064912,
  \href{http://dx.doi.org/10.1103/PhysRevC.80.064912}{\doi{10.1103/PhysRevC.80.064912}},
\href{http://www.arXiv.org/abs/0909.0191}{\texttt{arXiv:0909.0191}}.

\bibitem{Alver:2008gk}
\hrefCMSnoop {}{{PHOBOS} Collaboration, ``{System size dependence of cluster
  properties from two- particle angular correlations in Cu+Cu and Au+Au
  collisions at \rootsNN\ = 200\GeV}'',} \textit{ Phys. Rev. C} \textbf{ 81}
  (2010) 024904,
  \href{http://dx.doi.org/10.1103/PhysRevC.81.024904}{\doi{10.1103/PhysRevC.81.024904}},
\href{http://www.arXiv.org/abs/0812.1172}{\texttt{arXiv:0812.1172}}.

\bibitem{Alver:2009id}
\hrefCMSnoop {}{{PHOBOS} Collaboration, ``{High transverse momentum triggered
  correlations over a large pseudorapidity acceptance in Au+Au collisions at
  \rootsNN\ = 200\GeV}'',} \textit{ Phys. Rev. Lett.} \textbf{ 104} (2010)
  062301,
  \href{http://dx.doi.org/10.1103/PhysRevLett.104.062301}{\doi{10.1103/PhysRevLett.104.062301}},
\href{http://www.arXiv.org/abs/0903.2811}{\texttt{arXiv:0903.2811}}.

\bibitem{Abelev:2009jv}
\hrefCMSnoop {}{{STAR} Collaboration, ``{Three-particle coincidence of the long
  range pseudorapidity correlation in high energy nucleus-nucleus
  collisions}'',} \textit{ Phys. Rev. Lett.} \textbf{ 105} (2010) 022301,
  \href{http://dx.doi.org/10.1103/PhysRevLett.105.022301}{\doi{10.1103/PhysRevLett.105.022301}},
\href{http://www.arXiv.org/abs/0912.3977}{\texttt{arXiv:0912.3977}}.

\bibitem{Chatrchyan:2011eka}
\hrefCMSnoop {}{{CMS Collaboration}, ``{Long-range and short-range dihadron
  angular correlations in central PbPb collisions at a nucleon-nucleon center
  of mass energy of 2.76 TeV}'',} \textit{ JHEP} \textbf{ 07} (2011) 076,
  \href{http://dx.doi.org/10.1007/JHEP07(2011)076}{\doi{10.1007/JHEP07(2011)076}},
\href{http://www.arXiv.org/abs/1105.2438}{\texttt{arXiv:1105.2438}}.

\bibitem{Chatrchyan:2012wg}
\hrefCMSnoop {}{{CMS Collaboration}, ``{Centrality dependence of dihadron
  correlations and azimuthal anisotropy harmonics in PbPb collisions at
  \rootsNN\ = 2.76 TeV}'',} \textit{ Eur. Phys. J. C} \textbf{ 72} (2012) 2012,
  \href{http://dx.doi.org/10.1140/epjc/s10052-012-2012-3}{\doi{10.1140/epjc/s10052-012-2012-3}},
\href{http://www.arXiv.org/abs/1201.3158}{\texttt{arXiv:1201.3158}}.

\bibitem{Aamodt:2011by}
\hrefCMSnoop {}{{ALICE Collaboration}, ``{Harmonic decomposition of
  two-particle angular correlations in Pb-Pb collisions at \rootsNN\ =
  2.76\TeV}'',} \textit{ Phys. Lett. B} \textbf{ 708} (2012) 249,
  \href{http://dx.doi.org/10.1016/j.physletb.2012.01.060}{\doi{10.1016/j.physletb.2012.01.060}},
\href{http://www.arXiv.org/abs/1109.2501}{\texttt{arXiv:1109.2501}}.

\bibitem{ATLAS:2012at}
\hrefCMSnoop {}{{ATLAS Collaboration}, ``{Measurement of the azimuthal
  anisotropy for charged particle production in \rootsNN\ = 2.76 \TeV lead-lead
  collisions with the ATLAS detector}'',} \textit{ Phys. Rev. C} \textbf{ 86}
  (2012) 014907,
  \href{http://dx.doi.org/10.1103/PhysRevC.86.014907}{\doi{10.1103/PhysRevC.86.014907}},
\href{http://www.arXiv.org/abs/1203.3087}{\texttt{arXiv:1203.3087}}.

\bibitem{CMS:2013bza}
\hrefCMSnoop {}{{CMS Collaboration}, ``{Studies of azimuthal dihadron
  correlations in ultra-central PbPb collisions at \rootsNN\ = 2.76\TeV}'',}
  \textit{ JHEP} \textbf{ 02} (2014) 088,
  \href{http://dx.doi.org/10.1007/JHEP02(2014)088}{\doi{10.1007/JHEP02(2014)088}},
\href{http://www.arXiv.org/abs/1312.1845}{\texttt{arXiv:1312.1845}}.

\bibitem{Ollitrault:1992bk}
\hrefCMSnoop {}{J.-Y. Ollitrault, ``{Anisotropy as a signature of transverse
  collective flow}'',} \textit{ Phys. Rev. D} \textbf{ 46} (1992) 229,
\href{http://dx.doi.org/10.1103/PhysRevD.46.229}{\doi{10.1103/PhysRevD.46.229}}.

\bibitem{Alver:2010gr}
\hrefCMSnoop {}{B.~Alver and G.~Roland, ``Collision geometry fluctuations and
  triangular flow in heavy-ion collisions'',} \textit{ Phys. Rev. C} \textbf{
  81} (2010) 054905,
  \href{http://dx.doi.org/10.1103/PhysRevC.81.054905}{\doi{10.1103/PhysRevC.81.054905}},
  \href{http://www.arXiv.org/abs/1003.0194}{\texttt{arXiv:1003.0194}}.
Erratum: \DOI{10.1103/PhysRevC.82.039903}.

\bibitem{Voloshin:1994mz}
\hrefCMSnoop {}{S.~Voloshin and Y.~Zhang, ``{Flow study in relativistic nuclear
  collisions by Fourier expansion of azimuthal particle distributions}'',}
  \textit{ Z. Phys. C} \textbf{ 70} (1996) 665,
  \href{http://dx.doi.org/10.1007/s002880050141}{\doi{10.1007/s002880050141}},
\href{http://www.arXiv.org/abs/hep-ph/9407282}{\texttt{arXiv:hep-ph/9407282}}.

\bibitem{Alver:2010dn}
\hrefCMSnoop {}{B.~H. Alver, C.~Gombeaud, M.~Luzum, and J.-Y. Ollitrault,
  ``{Triangular flow in hydrodynamics and transport theory}'',} \textit{ Phys.
  Rev. C} \textbf{ 82} (2010) 034913,
  \href{http://dx.doi.org/10.1103/PhysRevC.82.034913}{\doi{10.1103/PhysRevC.82.034913}},
\href{http://www.arXiv.org/abs/1007.5469}{\texttt{arXiv:1007.5469}}.

\bibitem{Schenke:2010rr}
\hrefCMSnoop {}{B.~Schenke, S.~Jeon, and C.~Gale, ``Elliptic and triangular
  flow in event-by-event {D=3+1} viscous hydrodynamics'',} \textit{ Phys. Rev.
  Lett.} \textbf{ 106} (2011) 042301,
  \href{http://dx.doi.org/10.1103/PhysRevLett.106.042301}{\doi{10.1103/PhysRevLett.106.042301}},
\href{http://www.arXiv.org/abs/1009.3244}{\texttt{arXiv:1009.3244}}.

\bibitem{Qiu:2011hf}
\hrefCMSnoop {}{Z.~Qiu, C.~Shen, and U.~Heinz, ``{Hydrodynamic elliptic and
  triangular flow in Pb-Pb collisions at \rootsNN\ = 2.76\TeV}'',} \textit{
  Phys. Lett. B} \textbf{ 707} (2012) 151,
  \href{http://dx.doi.org/10.1016/j.physletb.2011.12.041}{\doi{10.1016/j.physletb.2011.12.041}},
\href{http://www.arXiv.org/abs/1110.3033}{\texttt{arXiv:1110.3033}}.

\bibitem{Aad:2015lwa}
\hrefCMSnoop {}{{ATLAS Collaboration}, ``{Measurement of the correlation
  between flow harmonics of different order in lead-lead collisions at
  \rootsNN\ = 2.76 TeV with the ATLAS detector}'',} \textit{ Phys. Rev. C}
  \textbf{ 92} (2015) 034903,
  \href{http://dx.doi.org/10.1103/PhysRevC.92.034903}{\doi{10.1103/PhysRevC.92.034903}},
\href{http://www.arXiv.org/abs/1504.01289}{\texttt{arXiv:1504.01289}}.

\bibitem{ALICE:2016kpq}
\hrefCMSnoop {}{{ALICE Collaboration}, ``{Correlated event-by-event
  fluctuations of flow harmonics in Pb-Pb collisions at \rootsNN\ = 2.76
  TeV}'',} \textit{ Phys. Rev. Lett.} \textbf{ 117} (2016) 182301,
  \href{http://dx.doi.org/10.1103/PhysRevLett.117.182301}{\doi{10.1103/PhysRevLett.117.182301}},
\href{http://www.arXiv.org/abs/1604.07663}{\texttt{arXiv:1604.07663}}.

\bibitem{Bilandzic:2013kga}
A.~Bilandzic\hrefCMSnoop {}{ {et~al.}, ``{Generic framework for anisotropic
  flow analyses with multiparticle azimuthal correlations}'',} \textit{ Phys.
  Rev. C} \textbf{ 89} (2014) 064904,
  \href{http://dx.doi.org/10.1103/PhysRevC.89.064904}{\doi{10.1103/PhysRevC.89.064904}},
\href{http://www.arXiv.org/abs/1312.3572}{\texttt{arXiv:1312.3572}}.

\bibitem{PhysRevC.95.044902}
\hrefCMSnoop {}{X.~Zhu, Y.~Zhou, H.~Xu, and H.~Song, ``{Correlations of flow
  harmonics in 2.76A TeV Pb--Pb collisions}'',} \textit{ Phys. Rev. C} \textbf{
  95} (2017) 044902,
  \href{http://dx.doi.org/10.1103/PhysRevC.95.044902}{\doi{10.1103/PhysRevC.95.044902}},
\href{http://www.arXiv.org/abs/1608.05305}{\texttt{arXiv:1608.05305}}.

\bibitem{Giacalone:2016afq}
\hrefCMSnoop {}{G.~Giacalone, L.~Yan, J.~Noronha-Hostler, and J.-Y. Ollitrault,
  ``{Symmetric cumulants and event-plane correlations in Pb+Pb collisions}'',}
  \textit{ Phys. Rev. C} \textbf{ 94} (2016) 014906,
  \href{http://dx.doi.org/10.1103/PhysRevC.94.014906}{\doi{10.1103/PhysRevC.94.014906}},
\href{http://www.arXiv.org/abs/1605.08303}{\texttt{arXiv:1605.08303}}.

\bibitem{Khachatryan:2010gv}
\hrefCMSnoop {}{{CMS Collaboration}, ``Observation of long-range near-side
  angular correlations in proton-proton collisions at the {LHC}'',} \textit{
  JHEP} \textbf{ 09} (2010) 091,
  \href{http://dx.doi.org/10.1007/JHEP09(2010)091}{\doi{10.1007/JHEP09(2010)091}},
\href{http://www.arXiv.org/abs/1009.4122}{\texttt{arXiv:1009.4122}}.

\bibitem{Aad:2015gqa}
\hrefCMSnoop {}{{ATLAS Collaboration}, ``Observation of long-range elliptic
  azimuthal anisotropies in $\sqrt{s} =$ 13 and 2.76 {TeV} pp collisions with
  the {ATLAS} detector'',} \textit{ Phys. Rev. Lett.} \textbf{ 116} (2016)
  172301,
  \href{http://dx.doi.org/10.1103/PhysRevLett.116.172301}{\doi{10.1103/PhysRevLett.116.172301}},
\href{http://www.arXiv.org/abs/1509.04776}{\texttt{arXiv:1509.04776}}.

\bibitem{Khachatryan:2015lva}
\hrefCMSnoop {}{{CMS Collaboration}, ``{Measurement of long-range near-side
  two-particle angular correlations in pp collisions at $\sqrt s =$ 13 TeV}'',}
  \textit{ Phys. Rev. Lett.} \textbf{ 116} (2016) 172302,
  \href{http://dx.doi.org/10.1103/PhysRevLett.116.172302}{\doi{10.1103/PhysRevLett.116.172302}},
\href{http://www.arXiv.org/abs/1510.03068}{\texttt{arXiv:1510.03068}}.

\bibitem{Khachatryan:2016txc}
\hrefCMSnoop {}{{CMS Collaboration}, ``{Evidence for collectivity in pp
  collisions at the LHC}'',} \textit{ Phys. Lett. B} \textbf{ 765} (2017) 193,
  \href{http://dx.doi.org/10.1016/j.physletb.2016.12.009}{\doi{10.1016/j.physletb.2016.12.009}},
\href{http://www.arXiv.org/abs/1606.06198}{\texttt{arXiv:1606.06198}}.

\bibitem{CMS:2012qk}
\hrefCMSnoop {}{{CMS Collaboration}, ``{Observation of long-range near-side
  angular correlations in proton-lead collisions at the LHC}'',} \textit{ Phys.
  Lett. B} \textbf{ 718} (2013) 795,
  \href{http://dx.doi.org/10.1016/j.physletb.2012.11.025}{\doi{10.1016/j.physletb.2012.11.025}},
\href{http://www.arXiv.org/abs/1210.5482}{\texttt{arXiv:1210.5482}}.

\bibitem{alice:2012qe}
\hrefCMSnoop {}{{ALICE Collaboration}, ``{Long-range angular correlations on
  the near and away side in \pPb\ collisions at \rootsNN\ = 5.02\TeV }'',}
  \textit{ Phys. Lett. B} \textbf{ 719} (2013) 29,
  \href{http://dx.doi.org/10.1016/j.physletb.2013.01.012}{\doi{10.1016/j.physletb.2013.01.012}},
\href{http://www.arXiv.org/abs/1212.2001}{\texttt{arXiv:1212.2001}}.

\bibitem{atlas:2012fa}
\hrefCMSnoop {}{{ATLAS Collaboration}, ``{Observation of Associated Near-side
  and Away-side Long-range Correlations in \rootsNN\ = 5.02\TeV Proton-lead
  Collisions with the ATLAS Detector}'',} \textit{ Phys. Rev. Lett.} \textbf{
  110} (2013) 182302,
  \href{http://dx.doi.org/10.1103/PhysRevLett.110.182302}{\doi{10.1103/PhysRevLett.110.182302}},
\href{http://www.arXiv.org/abs/1212.5198}{\texttt{arXiv:1212.5198}}.

\bibitem{Aaij:2015qcq}
\hrefCMSnoop {}{{LHCb Collaboration}, ``{Measurements of long-range near-side
  angular correlations in \rootsNN\ = 5 TeV proton-lead collisions in the
  forward region}'',} \textit{ Phys. Lett. B} \textbf{ 762} (2016) 473,
  \href{http://dx.doi.org/10.1016/j.physletb.2016.09.064}{\doi{10.1016/j.physletb.2016.09.064}},
\href{http://www.arXiv.org/abs/1512.00439}{\texttt{arXiv:1512.00439}}.

\bibitem{Khachatryan:2014jra}
\hrefCMSnoop {}{{CMS Collaboration}, ``{Long-range two-particle correlations of
  strange hadrons with charged particles in pPb and PbPb collisions at LHC
  energies}'',} \textit{ Phys. Lett. B} \textbf{ 742} (2015) 200,
  \href{http://dx.doi.org/10.1016/j.physletb.2015.01.034}{\doi{10.1016/j.physletb.2015.01.034}},
\href{http://www.arXiv.org/abs/1409.3392}{\texttt{arXiv:1409.3392}}.

\bibitem{ABELEV:2013wsa}
\hrefCMSnoop {}{{ALICE Collaboration}, ``{Long-range angular correlations of
  $\pi$, K and p in p--Pb collisions at \rootsNN\ = 5.02\TeV}'',} \textit{
  Phys. Lett. B} \textbf{ 726} (2013) 164,
  \href{http://dx.doi.org/10.1016/j.physletb.2013.08.024}{\doi{10.1016/j.physletb.2013.08.024}},
\href{http://www.arXiv.org/abs/1307.3237}{\texttt{arXiv:1307.3237}}.

\bibitem{Chatrchyan:2013nka}
\hrefCMSnoop {}{{CMS} Collaboration, ``{Multiplicity and transverse momentum
  dependence of two- and four-particle correlations in \pPb\ and \PbPb\
  collisions}'',} \textit{ Phys. Lett. B} \textbf{ 724} (2013) 213,
  \href{http://dx.doi.org/10.1016/j.physletb.2013.06.028}{\doi{10.1016/j.physletb.2013.06.028}},
\href{http://www.arXiv.org/abs/1305.0609}{\texttt{arXiv:1305.0609}}.

\bibitem{Aad:2014lta}
\hrefCMSnoop {}{{ATLAS Collaboration}, ``{Measurement of long-range
  pseudorapidity correlations and azimuthal harmonics in \rootsNN\ = 5.02 TeV
  proton-lead collisions with the ATLAS detector}'',} \textit{ Phys. Rev. C}
  \textbf{ 90} (2014) 044906,
  \href{http://dx.doi.org/10.1103/PhysRevC.90.044906}{\doi{10.1103/PhysRevC.90.044906}},
\href{http://www.arXiv.org/abs/1409.1792}{\texttt{arXiv:1409.1792}}.

\bibitem{Khachatryan:2015waa}
\hrefCMSnoop {}{{CMS Collaboration}, ``{Evidence for collective multi-particle
  correlations in pPb collisions}'',} \textit{ Phys. Rev. Lett.} \textbf{ 115}
  (2015) 012301,
  \href{http://dx.doi.org/10.1103/PhysRevLett.115.012301}{\doi{10.1103/PhysRevLett.115.012301}},
\href{http://www.arXiv.org/abs/1502.05382}{\texttt{arXiv:1502.05382}}.

\bibitem{Dusling:2015gta}
\hrefCMSnoop {}{K.~Dusling, W.~Li, and B.~Schenke, ``{Novel collective
  phenomena in high-energy proton-proton and proton-nucleus collisions}'',}
  \textit{ Int. J. Mod. Phys. E} \textbf{ 25} (2016) 1630002,
  \href{http://dx.doi.org/10.1142/S0218301316300022}{\doi{10.1142/S0218301316300022}},
\href{http://www.arXiv.org/abs/1509.07939}{\texttt{arXiv:1509.07939}}.

\bibitem{Schlichting:2014ipa}
\hrefCMSnoop {}{S.~Schlichting and B.~Schenke, ``{The shape of the proton at
  high energies}'',} \textit{ Phys. Lett. B} \textbf{ 739} (2014) 313,
  \href{http://dx.doi.org/10.1016/j.physletb.2014.10.068}{\doi{10.1016/j.physletb.2014.10.068}},
\href{http://www.arXiv.org/abs/1407.8458}{\texttt{arXiv:1407.8458}}.

\bibitem{Bozek:2016kpf}
\hrefCMSnoop {}{P.~Bo{\.{z}}ek, W.~Broniowski, and M.~Rybczy{\'n}ski,
  ``{Wounded quarks in A+A, p+A, and p+p collisions}'',} \textit{ Phys. Rev. C}
  \textbf{ 94} (2016) 014902,
  \href{http://dx.doi.org/10.1103/PhysRevC.94.014902}{\doi{10.1103/PhysRevC.94.014902}},
\href{http://www.arXiv.org/abs/1604.07697}{\texttt{arXiv:1604.07697}}.

\bibitem{Welsh:2016siu}
\hrefCMSnoop {}{K.~Welsh, J.~Singer, and U.~W. Heinz, ``{Initial state
  fluctuations in collisions between light and heavy ions}'',} \textit{ Phys.
  Rev. C} \textbf{ 94} (2016) 024919,
  \href{http://dx.doi.org/10.1103/PhysRevC.94.024919}{\doi{10.1103/PhysRevC.94.024919}},
\href{http://www.arXiv.org/abs/1605.09418}{\texttt{arXiv:1605.09418}}.

\bibitem{Chatrchyan:2014fea}
\hrefCMSnoop {}{{CMS Collaboration}, ``{Description and performance of track
  and primary-vertex reconstruction with the CMS tracker}'',} \textit{ JINST}
  \textbf{ 9} (2014) P10009,
  \href{http://dx.doi.org/10.1088/1748-0221/9/10/P10009}{\doi{10.1088/1748-0221/9/10/P10009}},
\href{http://www.arXiv.org/abs/1405.6569}{\texttt{arXiv:1405.6569}}.

\bibitem{Chatrchyan:2008zzk}
\hrefCMSnoop {}{{CMS Collaboration}, ``The {CMS} experiment at the {CERN}
  {LHC}'',} \textit{ JINST} \textbf{ 3} (2008) S08004,
\href{http://dx.doi.org/10.1088/1748-0221/3/08/S08004}{\doi{10.1088/1748-0221/3/08/S08004}}.

\bibitem{GEANT4}
\hrefCMSnoop {}{{GEANT4} Collaboration, ``{Geant4} --- a simulation toolkit'',}
  \textit{ Nucl. Instrum. Meth. A} \textbf{ 506} (2003) 250,
\href{http://dx.doi.org/10.1016/S0168-9002(03)01368-8}{\doi{10.1016/S0168-9002(03)01368-8}}.

\bibitem{Miller:2007ri}
\hrefCMSnoop {}{M.~L. Miller, K.~Reygers, S.~J. Sanders, and P.~Steinberg,
  ``{Glauber modeling in high energy nuclear collisions}'',} \textit{ Ann. Rev.
  Nucl. Part. Sci.} \textbf{ 57} (2007) 205,
  \href{http://dx.doi.org/10.1146/annurev.nucl.57.090506.123020}{\doi{10.1146/annurev.nucl.57.090506.123020}},
\href{http://www.arXiv.org/abs/nucl-ex/0701025}{\texttt{arXiv:nucl-ex/0701025}}.

\bibitem{Khachatryan:2016got}
\hrefCMSnoop {}{{CMS Collaboration}, ``{Observation of charge-dependent
  azimuthal correlations in pPb collisions and its implication for the search
  for the chiral magnetic effect}'',} \textit{ Phys. Rev. Lett.} \textbf{ 118}
  (2017) 122301,
  \href{http://dx.doi.org/10.1103/PhysRevLett.118.122301}{\doi{10.1103/PhysRevLett.118.122301}},
\href{http://www.arXiv.org/abs/1610.00263}{\texttt{arXiv:1610.00263}}.

\bibitem{Bilandzic:2010jr}
\hrefCMSnoop {}{A.~Bilandzic, R.~Snellings, and S.~Voloshin, ``{Flow analysis
  with cumulants: Direct calculations}'',} \textit{ Phys. Rev. C} \textbf{ 83}
  (2011) 044913,
  \href{http://dx.doi.org/10.1103/PhysRevC.83.044913}{\doi{10.1103/PhysRevC.83.044913}},
\href{http://www.arXiv.org/abs/1010.0233}{\texttt{arXiv:1010.0233}}.

\bibitem{Dusling:2017dqg}
\hrefCMSnoop {}{K.~Dusling, M.~Mace, and R.~Venugopalan, ``{Multiparticle
  collectivity from initial state correlations in high energy proton-nucleus
  collisions}'',} \textit{ Phys. Rev. Lett.} \textbf{ 120} (2018) 042002,
  \href{http://dx.doi.org/10.1103/PhysRevLett.120.042002}{\doi{10.1103/PhysRevLett.120.042002}},
\href{http://www.arXiv.org/abs/1705.00745}{\texttt{arXiv:1705.00745}}.

\bibitem{Dusling:2017aot}
\hrefCMSnoop {}{K.~Dusling, M.~Mace, and R.~Venugopalan, ``{Parton model
  description of multiparticle azimuthal correlations in $pA$ collisions}'',}
  \textit{ Phys. Rev. D} \textbf{ 97} (2018) 016014,
  \href{http://dx.doi.org/10.1103/PhysRevD.97.016014}{\doi{10.1103/PhysRevD.97.016014}},
\href{http://www.arXiv.org/abs/1706.06260}{\texttt{arXiv:1706.06260}}.

\bibitem{DiFrancesco:2016srj}
\hrefCMSnoop {}{P.~Di~Francesco, M.~Guilbaud, M.~Luzum, and J.-Y. Ollitrault,
  ``{Systematic procedure for analyzing cumulants at any order}'',} \textit{
  Phys. Rev. C} \textbf{ 95} (2017) 044911,
  \href{http://dx.doi.org/10.1103/PhysRevC.95.044911}{\doi{10.1103/PhysRevC.95.044911}},
\href{http://www.arXiv.org/abs/1612.05634}{\texttt{arXiv:1612.05634}}.

\bibitem{Jia:2017hbm}
\hrefCMSnoop {}{J.~Jia, M.~Zhou, and A.~Trzupek, ``{Revealing long-range
  multiparticle collectivity in small collision systems via subevent
  cumulants}'',} \textit{ Phys. Rev. C} \textbf{ 96} (2017) 034906,
  \href{http://dx.doi.org/10.1103/PhysRevC.96.034906}{\doi{10.1103/PhysRevC.96.034906}},
\href{http://www.arXiv.org/abs/1701.03830}{\texttt{arXiv:1701.03830}}.

\bibitem{Khachatryan:2015oea}
\hrefCMSnoop {}{{CMS Collaboration}, ``{Evidence for transverse momentum and
  pseudorapidity dependent event plane fluctuations in PbPb and pPb
  collisions}'',} \textit{ Phys. Rev. C} \textbf{ 92} (2015) 034911,
  \href{http://dx.doi.org/10.1103/PhysRevC.92.034911}{\doi{10.1103/PhysRevC.92.034911}},
\href{http://www.arXiv.org/abs/1503.01692}{\texttt{arXiv:1503.01692}}.

\end{thebibliography}\endgroup
